\documentclass[pra,twocolumn,showpacs,preprintnumbers,amsmath,amssymb]{revtex4} 
\usepackage{graphicx}
\usepackage{bm}
\usepackage{amsmath, amssymb}

\usepackage{color}

\begin{document}

\title{Spatially-modulated Superfluid States \\
in Fermionic Optical Ladder Systems with Repulsive Interactions}
\author{Yusuke Fujihara, Akihisa Koga and Norio Kawakami}
\affiliation{Department of Physics, Kyoto University, Kyoto, 606-8502, Japan}

\begin{abstract}
We investigate two-component ultracold fermionic atoms with repulsive interactions trapped in an optical lattice with a ladder structure. By applying the Bogoliubov-de Gennes equations to an effective {\it t-J} model in the strong correlation limit, we discuss how the spatially-modulated spin-singlet pairs with $d$-wave like symmetry are formed in the systems with trapping potentials. Furthermore, a close examination of the condensation energy as well as the local average of potential, kinetic and exchange energies by means of the variational Monte Carlo method elucidates that local particle correlations enhance the stability of the superfluid state via substantial energy gain due to singlet pairing in the high particle density region.
\end{abstract}

\pacs{03.75.Ss, 05.30.Fk, 74.72.-h}

\maketitle

\section{Introduction}

Ultracold atomic systems have attracted much interest since the successful realization of superfluidity in alkali metal atoms \cite{Anderson, Bradley, Davis, Greiner, Stoferle, Folling}. Among a number of interesting issues addressed in this field, fermionic atoms in an optical lattice \cite{Roati, Stoferle2}, formed by loading the ultracold atoms in a periodic potential, currently provide an extremely hot topic, where the superfluid \cite{Chin} and the Mott insulating states \cite{Jordens} as well as the Fermi surface \cite{Kohl} have been observed  recently. Extensive theoretical studies on attractive fermions in optical lattices have treated a wide variety of remarkable phenomena, such as the BEC-BCS crossover \cite{Koetsier, Zhao, Tamaki}, the superfluid-insulator transition \cite{Orso, Zhai, Moon, Iskin2, Chien}, and the possibility of a supersolid state \cite{Pour, Xianlong, Dao, Koga, Burkov}. Furthermore,  Fulde-Ferrell-Larkin-Ovchinnikov (FFLO)-type superfluid states have  been suggested in systems with trapping potentials \cite{Moreo, Parish, Chen, Iskin, Rizzi, Koponen}.

Fermionic systems with repulsive interactions, which have also been studied extensively, provide an ideal playground where we could realize various intriguing phenomena expected for highly correlated electrons in condensed matter physics \cite{Bloch2, Jaksch2, Morsch, Bloch}. A remarkable advantage in atomic gases over condensed matter systems is that their experimental parameters are highly controllable. For example, the depth of the optical lattice potential is controlled  by the light intensity, which leads to the precise alternation of atom tunneling between adjacent sites. Furthermore, one can easily create low-dimensional systems by tuning the laser beam configuration with the use of interference effects. From a viewpoint of quantum simulators of correlated electron systems, it may be particularly interesting to study the superfluidity in two-dimensional (2D) or ladder-type optical lattices since they are directly related to unconventional superconductors observed in 2D high-$T_{c}$ cuprates \cite{Bednorz, Yanase, Levin} and also some oxides with a ladder structure \cite{Akimitsu}. Therefore, it is an important issue to search for an unconventional superfluid state in repulsive fermions loaded in optical lattices. In this connection, we note that there is a crucial difference between the optical lattices and the ordinary solid state systems, {\it i.e.} the presence of trapping potential in the former system. It is thus desirable to clarify the effect of trapping potential in order to discuss the superfluid states of repulsive fermions in optical lattices.

In this paper, we study the stability of a $d$-wave like superfluid state of ultracold fermionic atoms in an optical lattice with a ladder structure. This study is motivated by the recent successful realization of a superlattice with double-well structure, which should certainly stimulate a systematic study of superfluidity on ladder-type optical lattices in the near future. As mentioned above, it is important to take into account the effects of repulsive interactions in the presence of inhomogeneity due to a trapping potential. By using the standard mean-field technique, we first describe the spatially-modulated spin-singlet pairs in a trapping potential. We further employ the variational Monte Carlo (VMC) method  to take into account particle correlations more precisely \cite{Yokoyama, Fujihara}. It is clarified  how the superfluid state with $d$-wave like symmetry is stabilized in our trapped system with particular emphasis on the role played by local particle correlations.

This paper is organized as follows.
We introduce the model Hamiltonian in \S \ref{Sec.Model}, and investigate how the spatially-modulated singlet pairing with $d$-wave like symmetry is realized 
in \S \ref{Sec.BdG}. In \S \ref{Sec.VMC}, we use the VMC method to discuss the stability of the superfluid state in the presence of strong correlations.
A brief summary is given in the last section.

\section{Model Hamiltonian} \label{Sec.Model}
We study a superfluid state of ultracold fermionic atoms in an optical lattice.
Here we consider a two-component fermionic system with Fermi-Fermi mixture such as ${}^{171}$Yb and ${}^{173}$Yb \cite{Fukuhara}. Alternatively, we can deal with fermions with two accessible hyperfine levels such as $|F,m_{F} \rangle = |9/2, -9/2 \rangle, |9/2, -7/2 \rangle$ for $^{40}$K \cite{Kohl}, where $F$ and $m_{F}$ are the total atomic angular momentum and its magnetic quantum number, respectively. By specifying the two different fermions by pseudo-spin indices $\sigma=\uparrow,\downarrow$, we can describe the ultracold atoms in an optical lattice in terms of the ordinary Hubbard Hamiltonian with on-site interactions \cite{Jaksch, Hofstetter}.
Here we focus on the strong coupling regime, where doubly occupied states hardly appear in the ground-state configuration, and then the antiferromagnetic superexchange terms between two adjacent sites become important to determine the nature of the ground state.
 In this regime, an effective model, the so-called $t-J$ model, is more convenient to discuss the ground state properties. The model Hamiltonian thus reads,
\begin{eqnarray}
\mathcal{H}_{t-J} &=& \mathcal{H}_{0} + \mathcal{H}', \label{Ht-J} \\
\mathcal{H}_{0}   &=& -t \sum_{\langle i,j \rangle \sigma} \left( c_{i\sigma}^{\dagger}c_{j\sigma} + h.c. \right) + \sum_{i\sigma} \left( V_{i} - \mu \right) n_{i\sigma}, 
\label{H0} \\
\mathcal{H}'      &=& J \sum_{\langle i,j \rangle} \left( \mathbf{\hat{S}}_{i} \cdot \mathbf{\hat{S}}_{j} - \frac{1}{4} n_{i}n_{j} \right), \label{H'}
\end{eqnarray}
where $c_{i\sigma}^{\dagger}$ ($c_{i\sigma}$) is a creation (annihilation) operator of a fermion with spin $\sigma$ at site $i$, 
$n_{i\sigma} = c_{i\sigma}^{\dagger}c_{i\sigma}$
and $n_{i} = n_{i\uparrow} + n_{i\downarrow}$.
$\mathbf{\hat{S}}_{i} = (1/2)\sum_{\alpha,\beta}c_{i\alpha}^{\dagger}\boldsymbol{\hat{\sigma}}_{\alpha\beta}c_{i\beta}$, where $\boldsymbol{\hat{\sigma}}$ is the Pauli matrix.
$t$ and $J (>0)$ are the tunneling matrix and the antiferromagnetic exchange interaction between nearest neighbour sites, and $\mu$ the chemical potential.
The interacting fermions are trapped in the harmonic potential $V_{i}$.
Note that the doubly occupied states are excluded at each site in the Hamiltonian (\ref{Ht-J}) due to the strong onsite repulsive interactions.

\begin{figure}[t!]
\begin{tabular}{cc}
\hspace{-5mm}
\resizebox{50mm}{!}{\includegraphics{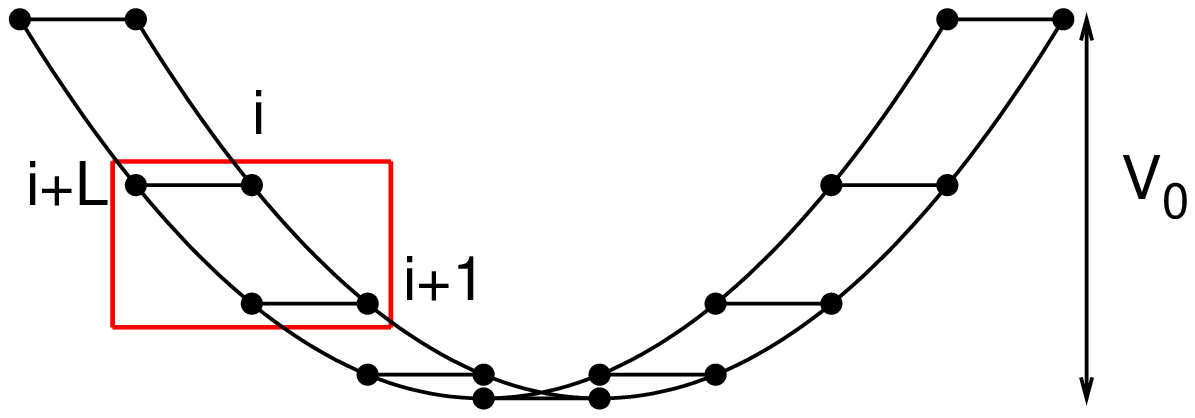}} &
\hspace{3mm}
\resizebox{25mm}{!}{\includegraphics{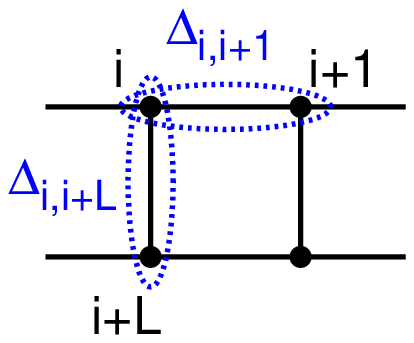}} \\
\end{tabular}
\caption{(Color online) Schematic pictures of a ladder structure ($L \times 2$ sites). 
$V_{0}$ is the depth of the harmonic potential in  the system.
$\Delta_{i,i+1}$ and $\Delta_{i,i+L}$ are the nearest neighbour pair potentials  along the leg and the rung directions (see text).}
\label{Fig_Model}
\end{figure}

In this paper, we deal with the optical lattice with a ladder structure schematically shown in Fig.\ref{Fig_Model}. Owing to the existence of the rung in the ladder, this quasi-1D system may have essence of 2D systems: namely a $d$-wave superfluid state may be realized in the strong correlation limit. It is important to note that this sort of ladder system can be experimentally realized by double well potentials \cite{Strabley, Danshita}.  
The tunneling matrix $t$ and the exchange coupling $J$ are given such that
$t \sim (4/\sqrt{\pi}) E_{R} (E_{R}/v_{0})^{3/4} \exp [-2 \sqrt{v_{0}/E_{R}}]$ 
and 
$J \sim t^{2}/U \sim \sqrt{32/\pi} (ka_{s})^{-1} E_{R} (E_{R}/v_{0})^{3/4} \exp [-4 \sqrt{v_{0}/E_{R}}]$,
within the harmonic approximation \cite{Hofstetter, Jaksch, Bloch}, where $a_{s}$ is the $s$-wave scattering length and $E_{R} = \hbar^{2}k^{2}/2m$ the atomic recoil energy in the optical lattice potential.
$m$, $v_{0}$ and $k$ are the mass of the atom, the intensity and the wave number of the lattice potential $V_{lat}(\mathbf{x}) = v_{0} \sum_{i} \sin(kx^{2}_{i})$.
Thus the exchange interaction $J \sim t^{2}/U$ can be experimentally controlled.
Furthermore, some elegant techniques to control $J$ have been proposed by means of the spin-dependent lattice potentials \cite{Duan, Kuklov, Ripoll} and the superlattice potentials \cite{Trotzky}.
Therefore, the ladder system has some advantages in studying the stability of the superfluid state with $d$-wave like symmetry.
In this study, we consider the ladder system with $L \times 2 (L=50)$ sites 
and fix the total number of atoms as $N_{\uparrow} = N_{\downarrow} = 25$, where $N_{\sigma} = \sum_{i}n_{i\sigma}$.
The harmonic trapping potential is given as $V_{i}=0$ at four sites around the center of the system and $V_{i}=4.0t$ at the edges. Note also that  $V_{i} = V_{L-1-i} = V_{i+L} = V_{2L-1-i}$. We fix the chemical potential $\mu$ at the Fermi level in the non-interacting case and use $t$ as an energy unit.

In the following, we discuss the possibility of a $d$-wave superfluid state in our optical ladder system. Since the fermionic atoms are confined in the inhomogeneous potential, it is not straightforward to deal with the effects of particle correlations. In this paper, we make use of the Bogoliubov-de Gennes (BdG) equations and the VMC method with a proper trial function to deal with this problem.

\section{Spatially-Modulated Singlet Pairing State} \label{Sec.BdG}
In this section, we discuss how the spatially-modulated superfluid state is realized by means of the BdG mean-field theory.
We here introduce site-dependent singlet pairing mean-fields
$\Delta_{ij} = \langle c_{i \downarrow} c_{j \uparrow} \rangle$ 
($ = \langle c_{j \downarrow} c_{i \uparrow} \rangle$).
In terms of the Bogoliubov transformation 
$c_{i \sigma} = \sum_{\lambda} \big\{ u_{i}^{\lambda} a_{\lambda \sigma} 
- \sigma v_{i}^{\lambda *} a_{\lambda \bar{\sigma}}^{\dagger} \big\}$,
we obtain the following BdG equations \cite{Ogata},
\begin{eqnarray}
\sum_{j} \left(
	\begin{array}{cc}
	H_{ij}     & F_{ij} \\
	F_{ji}^{*} & -H_{ji}^{*}\\
\end{array}
\right)
\left(
	\begin{array}{c}
	u_{j}^{\lambda} \\
	v_{j}^{\lambda *} \\
	\end{array}
\right)
= E_{\lambda} \left(
	\begin{array}{c}
	u_{i}^{\lambda} \\
	v_{i}^{\lambda *} \\
	\end{array}
\right),
\label{BdGeq}
\end{eqnarray}
with
\begin{eqnarray}
H_{ij} &=& -t \delta_{\langle ij \rangle} + \big( V_{i} - \mu \big) \delta_{ij}, \label{H_ij} \\
F_{ij} &=& -J \Delta_{ij} \delta_{\langle ij \rangle}, \label{F_ij}
\end{eqnarray}
and the self-consistent equations,
\begin{eqnarray}
\Delta_{ij} = - \sum_{\lambda} u_{i}^{\lambda} v_{j}^{\lambda *},
\label{SCeq}
\end{eqnarray}
where $\delta_{\langle ij \rangle}$ is the Kronecker delta for the nearest neighbour sites $i$ and $j$. We ignore the constraint for doubly occupied states in the Hamiltonian (\ref{Ht-J}) for a while, which largely reduces our numerical efforts. In the periodic case, it is known that the constraint mainly yields the renormalization of tunneling matrix $t$ \cite{Ogata}. In our inhomogeneous case, however, the situation is  more complicated. The constraint will be carefully treated in the next section.

\begin{figure}[t]
\begin{tabular}{cc}
\hspace{-2mm}
\resizebox{42mm}{!}{\includegraphics{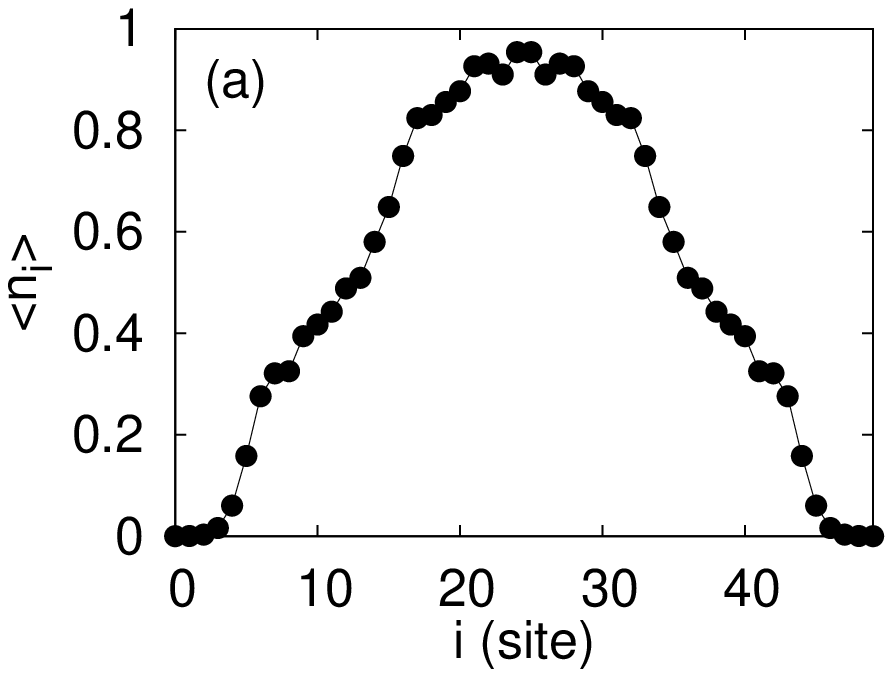}} &
\resizebox{42mm}{!}{\includegraphics{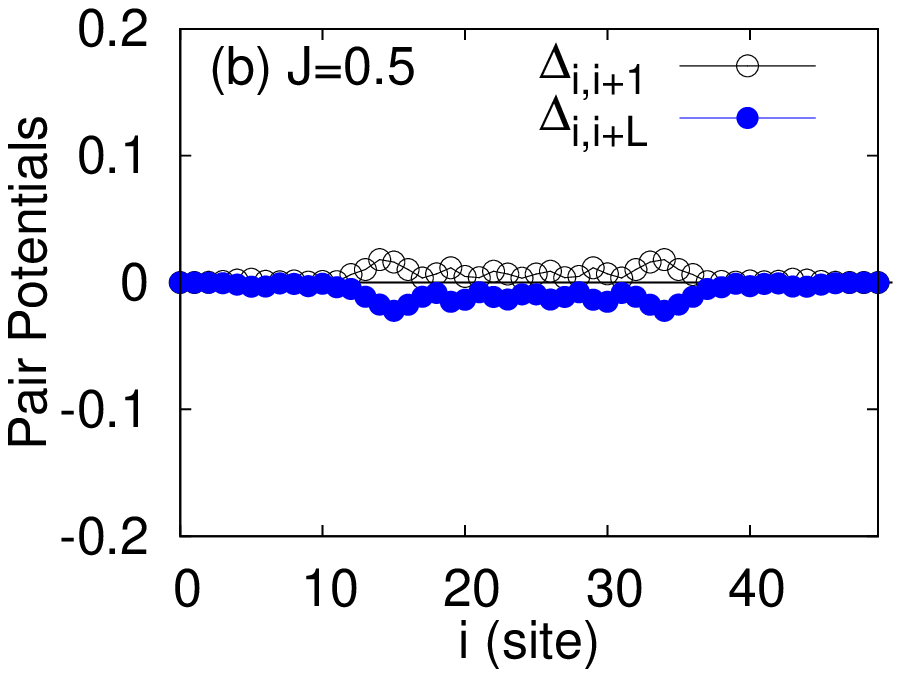}} \\
\hspace{-2mm}
\resizebox{42mm}{!}{\includegraphics{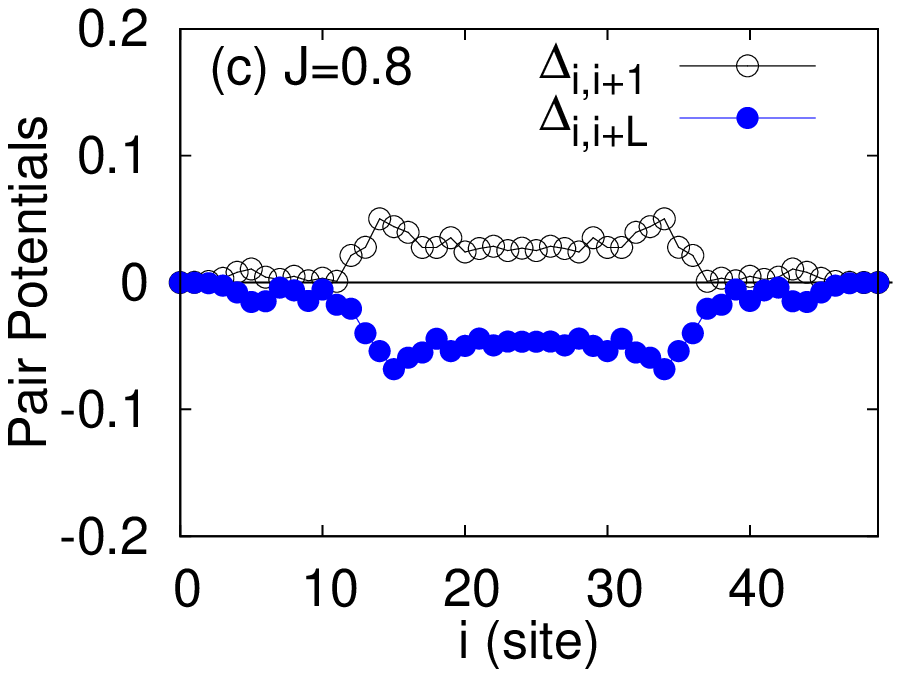}} &
\resizebox{42mm}{!}{\includegraphics{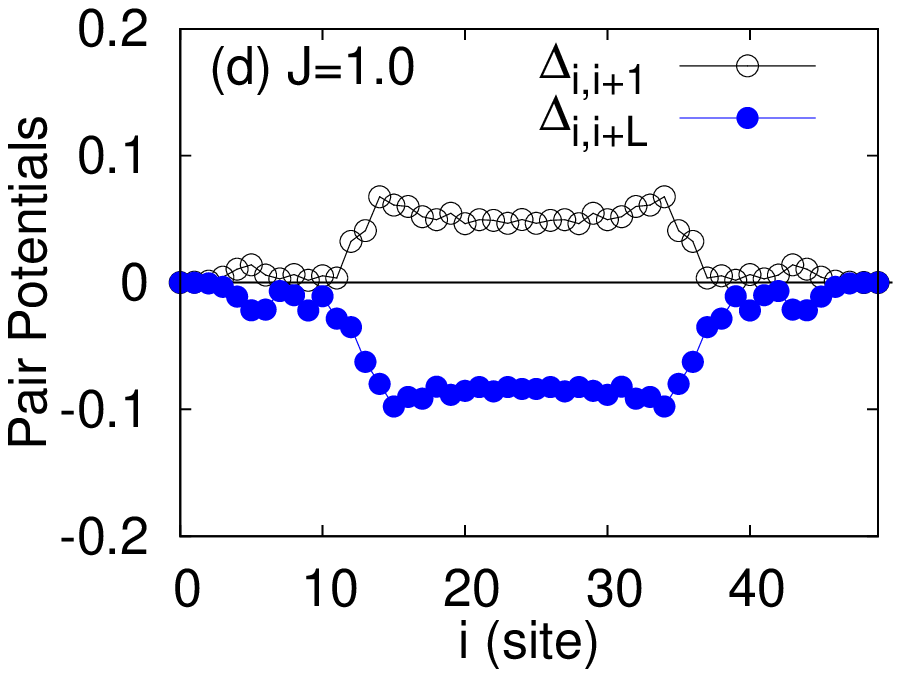}} \\
\hspace{-2mm}
\resizebox{42mm}{!}{\includegraphics{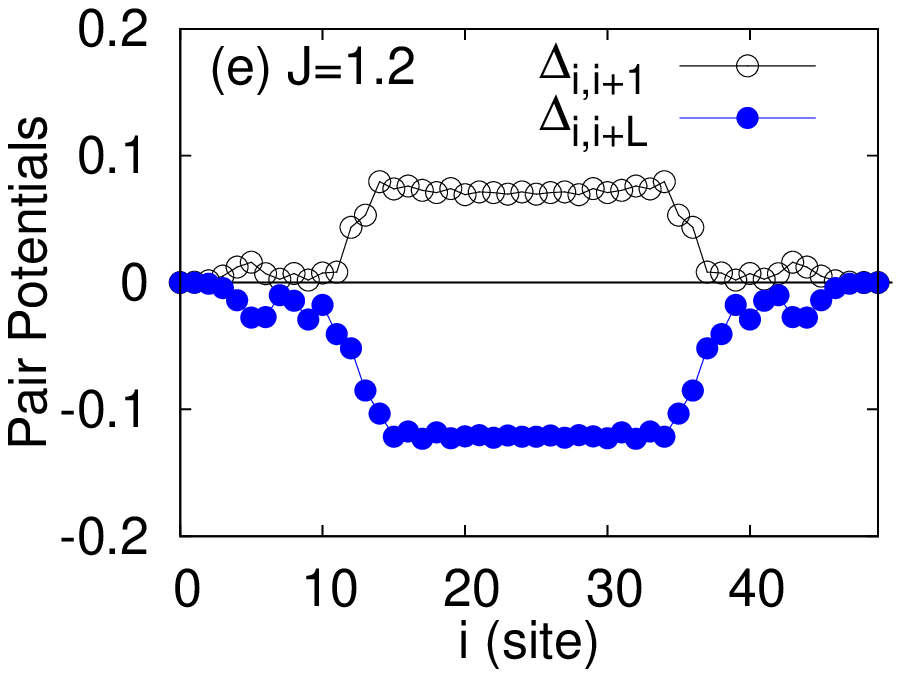}} &
\resizebox{42mm}{!}{\includegraphics{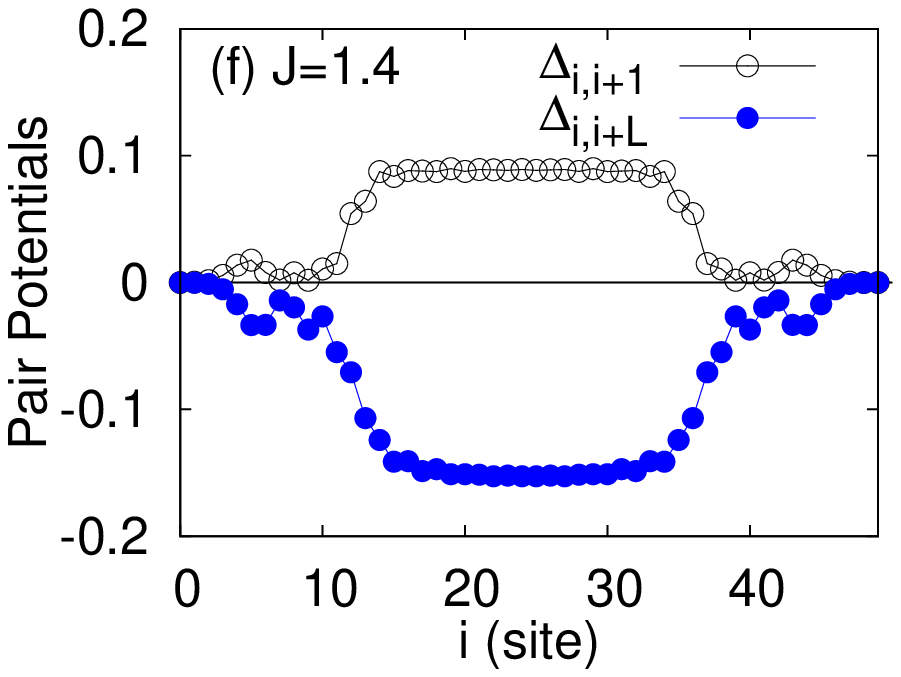}} \\
\end{tabular}
\caption{(Color online) (a) Spatial distributions of the particle density $\langle n_{i}\rangle (=\langle n_{i+L}\rangle)$, (b)-(f) the pair potentials along the leg direction $\Delta_{i,i+1}$ (open circles) and the rung direction $\Delta_{i,i+L}$ (filled circles) for $J=0.5, 0.8, 1.0, 1.2$ and $1.4$ (see Fig.\ref{Fig_Model}).}
\label{Fig_BdG}
\end{figure}

\begin{figure}[h]
\begin{center}
\includegraphics[width=7.0cm]{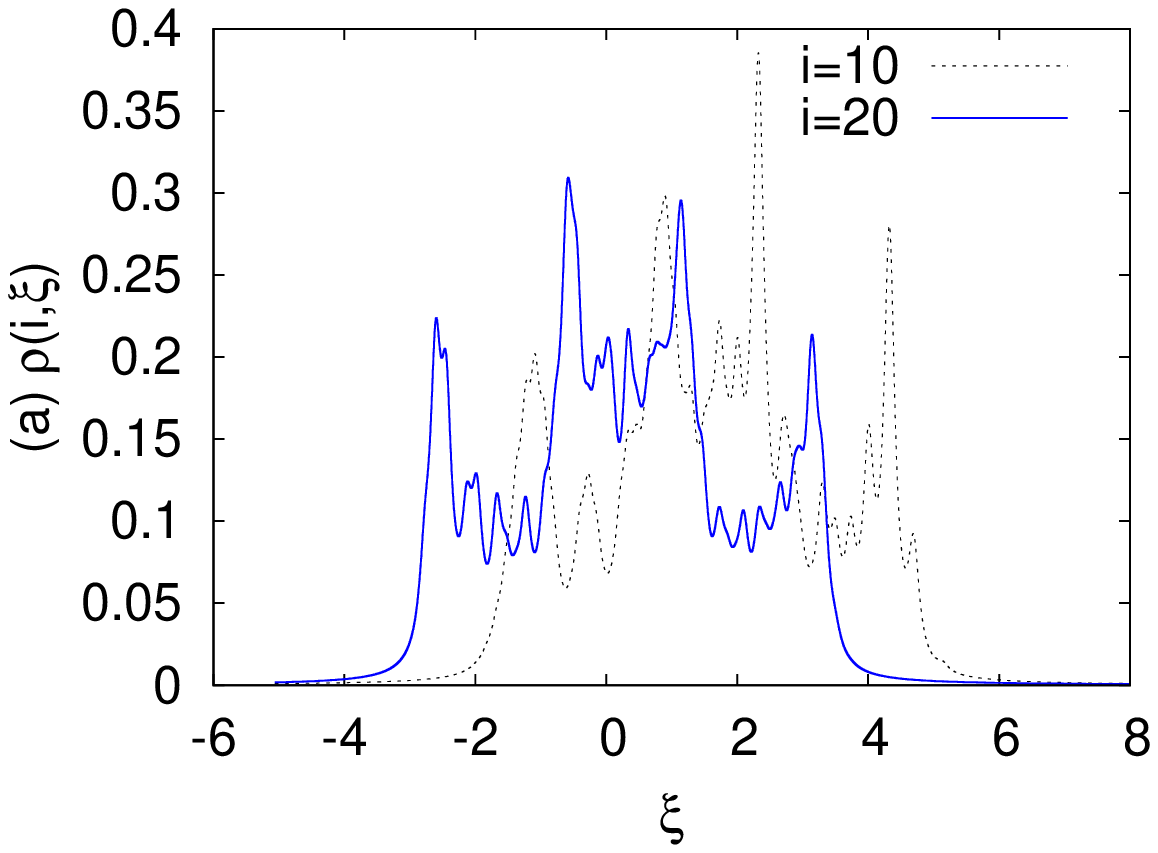} \\ \label{Fig_LDOS}
\includegraphics[width=7.0cm]{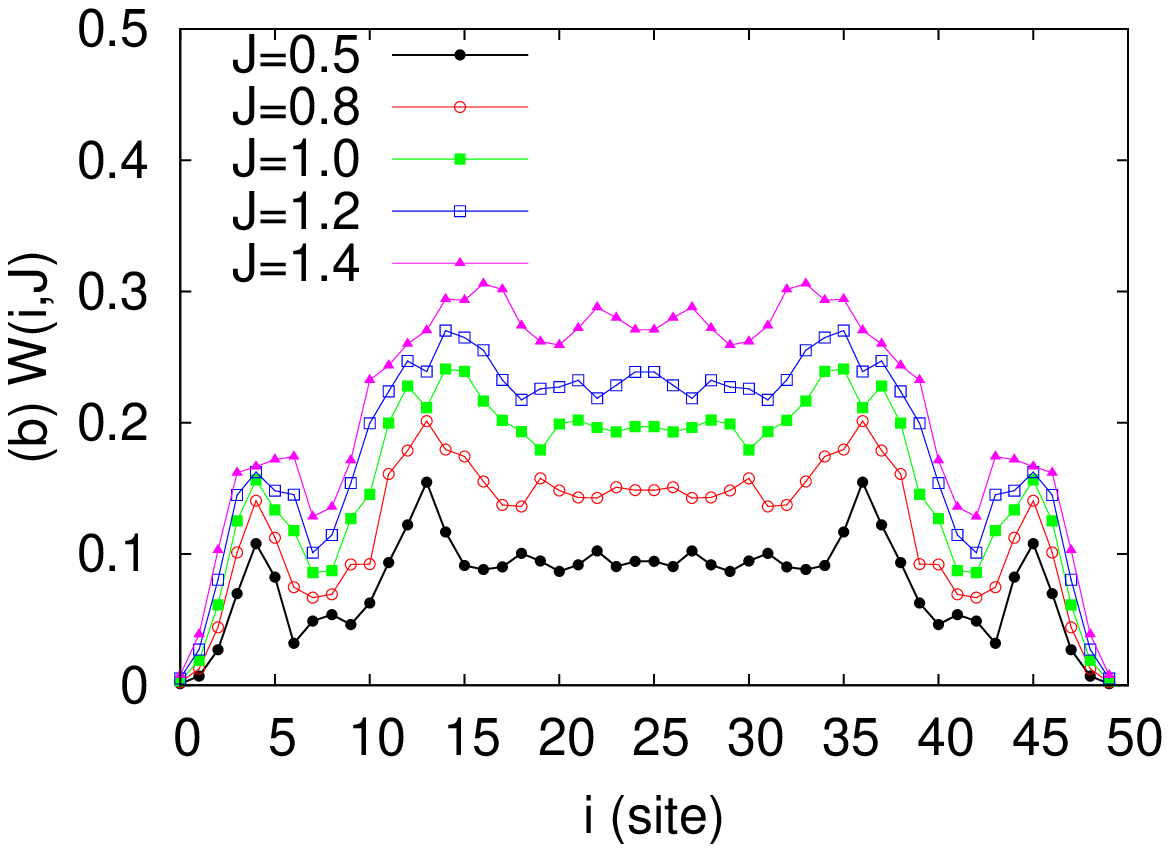} \label{Fig_W}
\caption{(Color online) (a) LDOS $\rho(i,\xi) = \sum_{n} |\phi_{i}^{n}|^{2} \delta (\xi - \xi_{n})$ at $i=10$ (dotted line) and $i=20$ (solid line) :delta functions are broadened with width of $\eta = 0.2t$.
(b) The integrated LDOS around the Fermi level $W(i,J)$ in eq.(\ref{W}) for $J=0.5$ (filled circles), $0.8$ (open circles), $1.0$ (filled squares), $1.2$ (open square) and $1.4$ (filled triangles).}
\end{center}
\end{figure}

We first calculate the site-dependent particle density, $\langle n_{i}\rangle  = \sum_{\lambda} |v_{i}^{\lambda}|^2$. The results are shown in Fig.\ref{Fig_BdG} (a). It is seen that the particles are smoothly distributed in the lattice. Its profile is mainly determined by the curvature of the harmonic potential, so that it is hardly affected by the exchange interaction $J$. On the other hand, the pair potential between the nearest neighbour sites $\Delta_{ij}$ strongly depends on the magnitude of the exchange interaction, as seen in Figs.\ref{Fig_BdG} (b)-(f). For $J<0.5$, the pair potential is too small to detect within our numerical accuracy, where a singlet pairing state is not realized, and the system is still in the normal ground state. On the other hand, for $J>0.5$, the pair potential takes finite values, as shown in Figs.\ref{Fig_BdG} (b)-(f), implying that the intersite exchange coupling stabilizes the singlet-pairing superfluid state.
Here, we find two characteristic features in the profile of the pair potential. One is the cusp (or peak) structures around $i=5$ and $i=15$, and the other is the plateau around the center. The above characteristic properties reflect atomic states inherent in our ladder system. To see this clearly, we introduce the integrated local density of states (LDOS), $W(i,J)$, around the Fermi level $\epsilon_{F}$. The quantity is defined as,
\begin{eqnarray}
W(i,J) &\equiv& \int_{\epsilon_{F}-\frac{J}{2}}^{\epsilon_{F}+\frac{J}{2}}\rho(i,\xi)d\xi, \nonumber \\
     &=     & \sum_{n} \int_{\epsilon_{F}-\frac{J}{2}}^{\epsilon_{F}+\frac{J}{2}} |\phi_{i}^{n}|^{2} \delta(\xi-\xi_{n}) d\xi,
\label{W}
\end{eqnarray}
where $\rho(i,\xi)$ denotes the LDOS at site $i$, and $\xi_{n}$ and  ${\bm \phi}^{n} = \{ \phi_{0}^{n}, \phi_{1}^{n}, \cdots \phi_{2L-1}^{n} \}$  are the eigenenergies and the eigenfunctions for the non-interacting Hamiltonian 
$\mathcal{H}_{0}$, respectively. Let us first take a look at the LDOS $\rho(i,\xi)$ for $\mathcal{H}_{0}$ shown in Fig.\ref{Fig_W} (a). It is seen that the shape of the LDOS at site $i$ reflects the DOS for a uniform ladder system that has a characteristic four-peak structure originating from the van-Hove edge singularity in one dimension. At first glance, it is not evident whether this LDOS is really related to the spatially modulated pair potential. However, the integrated LDOS over the range of $J$, $W(i,J)$,  makes this point clear. In Fig.\ref{Fig_W} (b), it is found that the cusp (or peak) structures indeed appear around $i = 5$ and $15$ in $W(i,J)$, and the plateau with some fluctuations is formed around the center. This implies that the spatial variation of the pair potentials reflect the integrated LDOS, in agreement with the fact that the particles around the Fermi level are relevant to the formation of spin-singlet pairs in the superfluid state. An important point to be noticed here is that the pair potentials along the leg and the rung directions have opposite signs as shown in Figs.\ref{Fig_BdG} (b)-(f), implying that the pairing state with $d_{x^{2}-y^{2}}$-like symmetry is stabilized in our inhomogeneous system. This is naturally expected from the $d$-wave superconducting state realized in the doped ladder system without the trapping potential \cite{Sigrist, Troyer, Orignac, Sorella}.

We have so far neglected the local constraint due to strong onsite repulsion at each site. Although this kind of approximation has been employed in some cases \cite{Ogata, Zhang, Ogata2, Ogata3}, it is not sufficient to discuss the stability of superfluidity in our correlated system: for example, the particles in the normal phase are completely free, and also the onsite pair potential $\Delta_{ii}$  can be finite in the large $J$ case, in contrast to the ordinary $d$-wave state in the homogeneous system. To remedy these pathological points, a careful treatment of local particle correlations is necessary, which will be done in the next section.

\section{VMC study of particle correlations} \label{Sec.VMC}

In this section, we perform the VMC calculation to elucidate how the local particle correlations affect the superfluid state. 
Let us introduce a trial state, 
\begin{eqnarray}
|\Psi \rangle = \mathcal{P}_{G} |\Phi_{BCS} (\{\Delta_{ij}\}, g)\rangle,
\label{TWF}
\end{eqnarray}
where $\mathcal{P}_{G} = \prod_{i} [1-n_{i\uparrow}n_{i\downarrow}]$ 
is a Gutzwiller projection operator \cite{Gutzwiller}, which completely excludes the doubly occupied states at each site.
$|\Phi_{BCS} (\{\Delta_{ij}\}, g)\rangle$ is a BCS state with variational parameters $(\{\Delta_{ij}\}, g)$ to describe the superfluid state with spatially-modulated pairing correlations. This state is explicitly given as, 
\begin{eqnarray}
|\Phi_{BCS} \rangle &=& \prod_{n} \left( 
u_{n}+v_{n}c_{n\uparrow}^{\dagger}c_{n\downarrow}^{\dagger} \right) 
|0\rangle, \nonumber \\
              &\propto& \left( \sum_{n} \frac{u_{n}}{v_{n}} 
c_{n\uparrow}^{\dagger} c_{n\downarrow}^{\dagger} \right)^{N/2} |0\rangle,
\label{BCS}
\end{eqnarray}
with
\begin{eqnarray}
u_{n}^{2} &=& \frac{1}{2} \left[ 1+\frac{\xi_{n}}{\sqrt{\xi_{n}^{2}+\Delta_{n}^{2}}} \right], \label{un} \\
v_{n}^{2} &=& \frac{1}{2} \left[ 1-\frac{\xi_{n}}{\sqrt{\xi_{n}^{2}+\Delta_{n}^{2}}} \right], \label{vn} 
\end{eqnarray}
where $c_{n\sigma}=\sum_{i} \phi_{i}^{n} c_{i\sigma}$ and $N = N_{\uparrow} + N_{\downarrow}$.
The gap function $\Delta_n$ is then given as,
\begin{eqnarray}
\Delta_{n} &=& -J \langle c_{n\downarrow}c_{n\uparrow}\rangle, \nonumber \\
&=& -g\sum_{ij}\phi_{i}^{n}\phi_{j}^{n} \Delta_{ij}, \label{Delta_n}
\end{eqnarray}
where $g$ is a variational parameter controlling the amplitude of the pair potential, while  $\Delta_{ij}$ determines its spatial variation according to the solution of BdG equations. By optimizing the variational parameters so as to minimize the total energy, we discuss the ground state properties in the optical lattice system. It is still not easy to determine these parameters since the number of parameters is very large in the presence of the trapping  potential. Recall here that in the solutions obtained by the BdG equations, the spatial dependence of $\Delta_{ij}$ is controlled by $J$. This observation enables us to approximately reduce the parameter space by regarding $J$ in eq. (\ref{F_ij}) as a variational parameter to control the distribution of $\{\Delta_{ij}\}$ obtained from the BdG equations.
This new variational parameter is denoted as $\tilde{J}$ in the following. 
In order to incorporate the strong correlation effects in the order parameter, we here set the onsite component of the pair potential to be zero, $\Delta_{ii} =0$, and  retain only the nearest neighbor component of pair potential, $\Delta_{ij} \rightarrow \Delta_{ij}\delta_{\langle ij\rangle}$, which gives the most significant contribution to the superfluid state. This kind of approximation should be justified in the strong coupling regime.  Note that the correlation effects in the normal state can be properly taken into account by the ordinary VMC procedure.

\begin{figure}[h]
\begin{tabular}{cc}
\hspace{-2mm}
\resizebox{45mm}{!}{\includegraphics{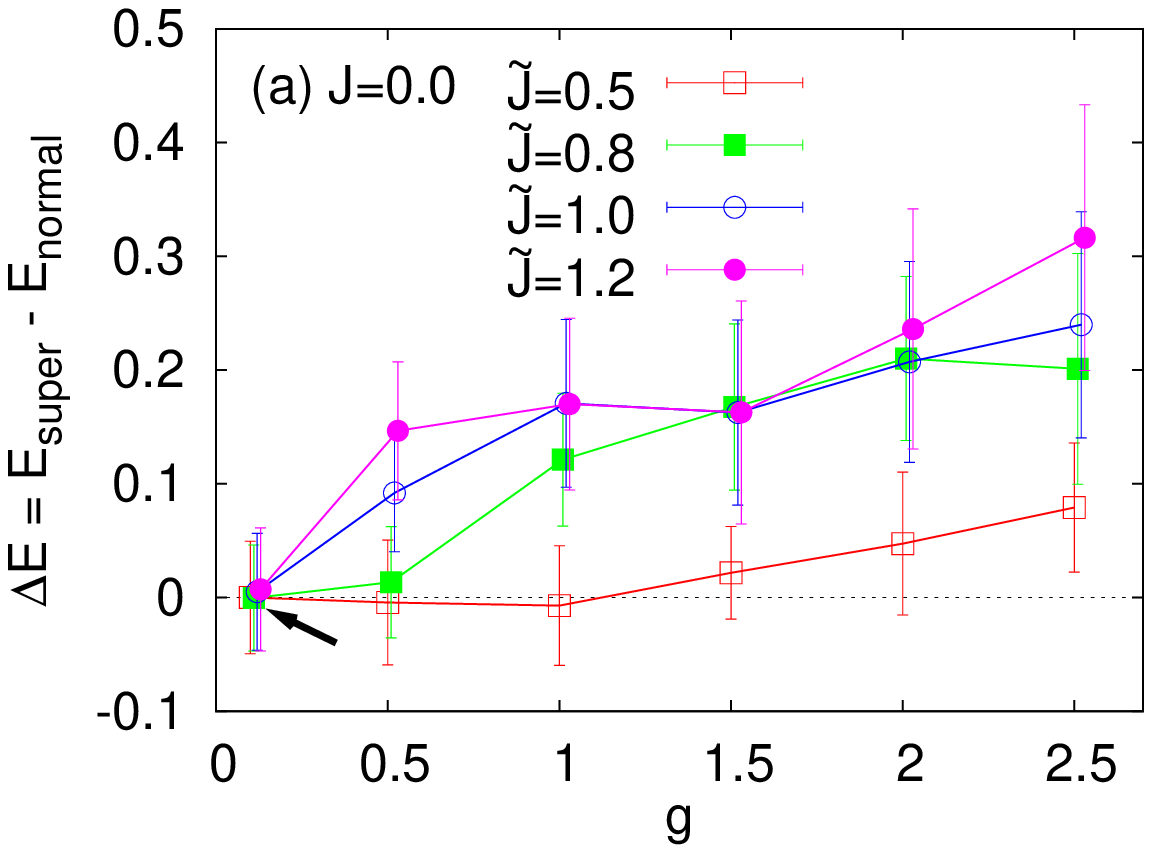}} &
\hspace{-4mm}
\resizebox{45mm}{!}{\includegraphics{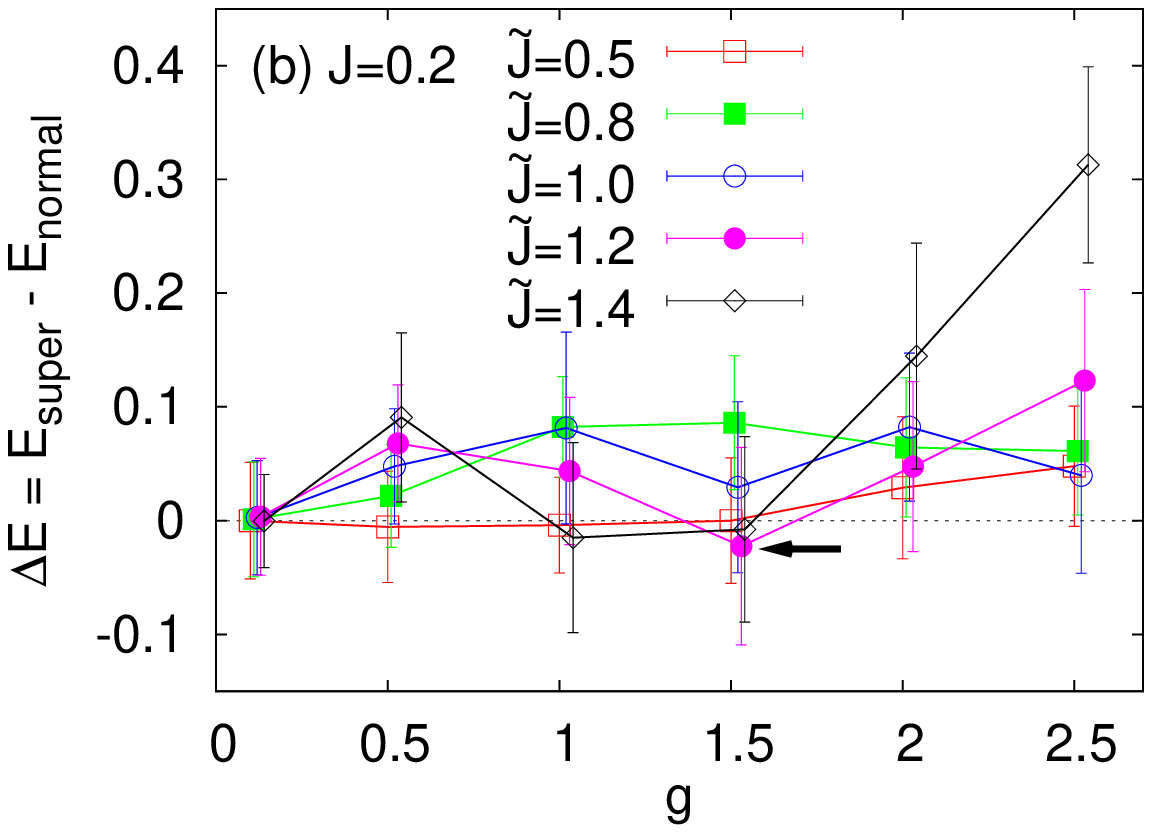}} \\
\hspace{-2mm}
\resizebox{45mm}{!}{\includegraphics{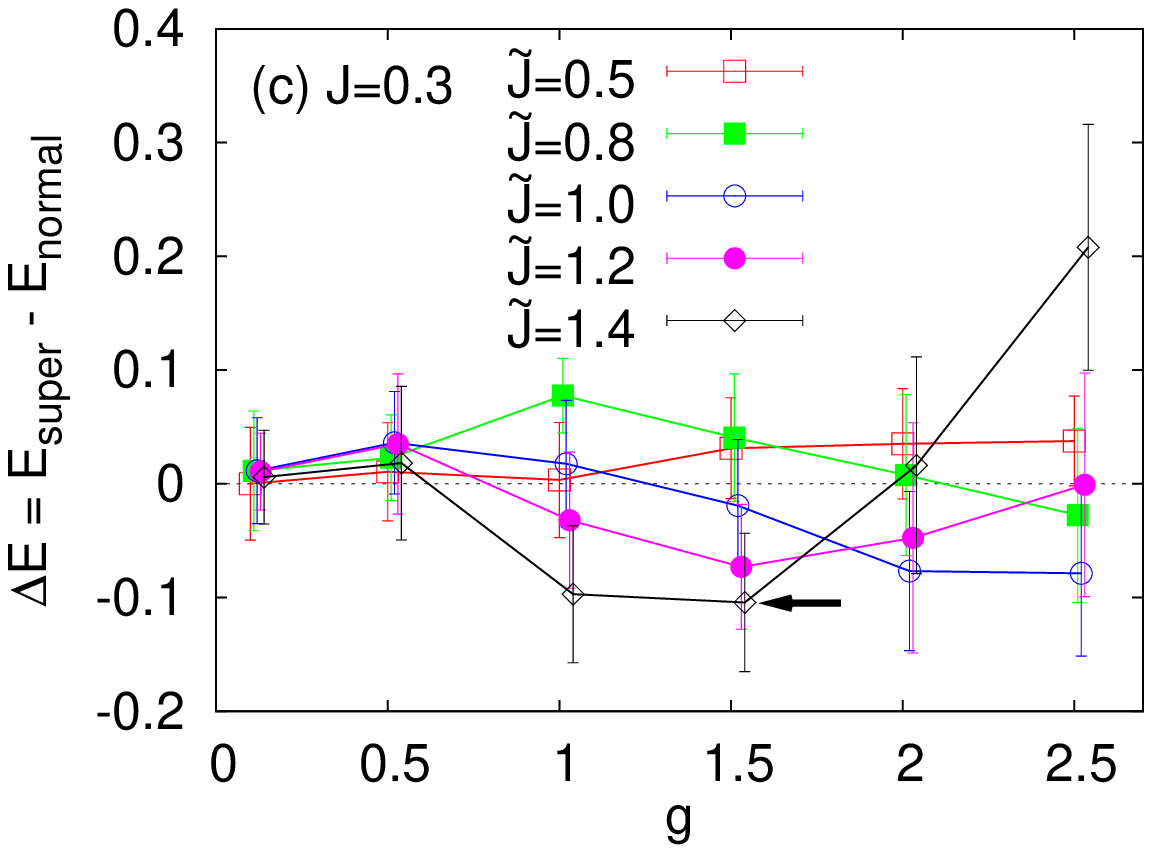}} &
\hspace{-4mm}
\resizebox{45mm}{!}{\includegraphics{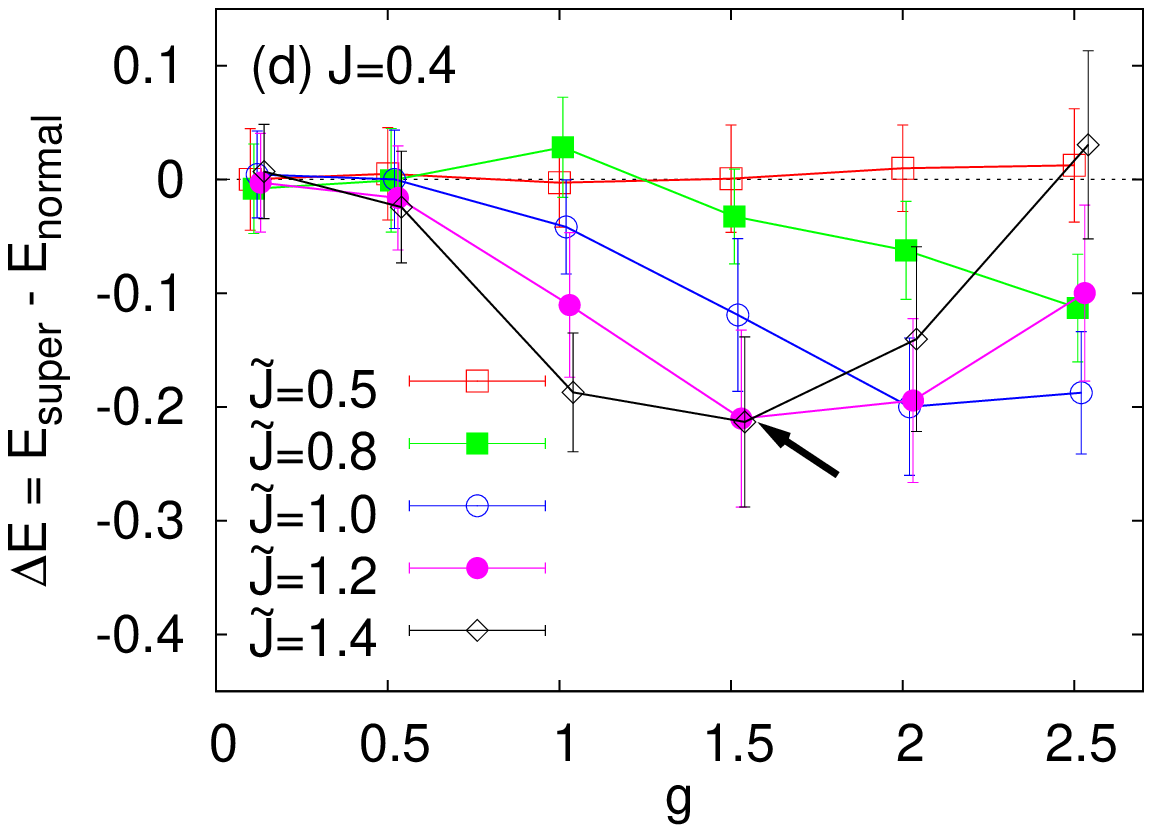}} \\
\end{tabular}
\caption{(Color online) Condensation energies as a function of the variational parameter $g$ for (a)$J=0$, (b)$0.2$, (c)$0.3$ and (d)$0.4$ obtained by the VMC method. The sets of $\{ \Delta_{ij} \}$ used in each calculation are obtained for  ${\tilde J}=0.5$ (open squares), $0.8$ (filled squares), $1.0$ (open circles), $1.2$ (filled circles) and $1.4$ (open diamonds), which are shown in Figs.\ref{Fig_BdG} (a)-(d). 
}
\label{Fig_dE}
\end{figure}

By performing the VMC simulations with samplings $\sim 10^{6}$, we obtain the condensation energy $\Delta E = E_{super} - E_{normal}$, for the variational parameters $(\tilde{J}, g)$ as shown in Fig.\ref{Fig_dE}, 
where $E_{super}$ and $E_{normal}$ are the energies of the superfluid state with $\Delta_{n}\neq 0$ and the normal state with $\Delta_{n}=0$ for all $n$.
At $J=0$, the model is reduced to the system only with correlated hopping. 
In the case, the formation of the singlet pairs gives rise to the loss of the kinetic energy. Therefore, the normal state is more stable than the superfluid state with any values of $\tilde{J}$, as shown in Fig.\ref{Fig_dE} (a).
On the other hand, the increase in the exchange coupling $J$ favours the superfluid state with singlet pairing correlations, where the normal state is no longer the ground state. In fact, it is found in Figs.\ref{Fig_dE} (b)-(d) that the condensation energy $\Delta E = E_{super} - E_{normal}$ has a minimum at a finite $g$ although statistical errors are not so small in our calculations. This suggests that the spatially modulated superfluid state with $d$-wave like symmetry is indeed stabilized by exchange interactions. It should be noted that the critical value of $J \sim 0.2$ is smaller than $J \sim 0.5$ estimated in the previous section without electron correlations, which implies that the correlation effects enhance the stability of superfluid state. At first glance, this conclusion seems a little bit unusual, since the correlation effects normally have a tendency to suppress the ordering.

\begin{figure}[b]
\begin{tabular}{cc}
\hspace{-2mm}
\resizebox{42mm}{!}{\includegraphics{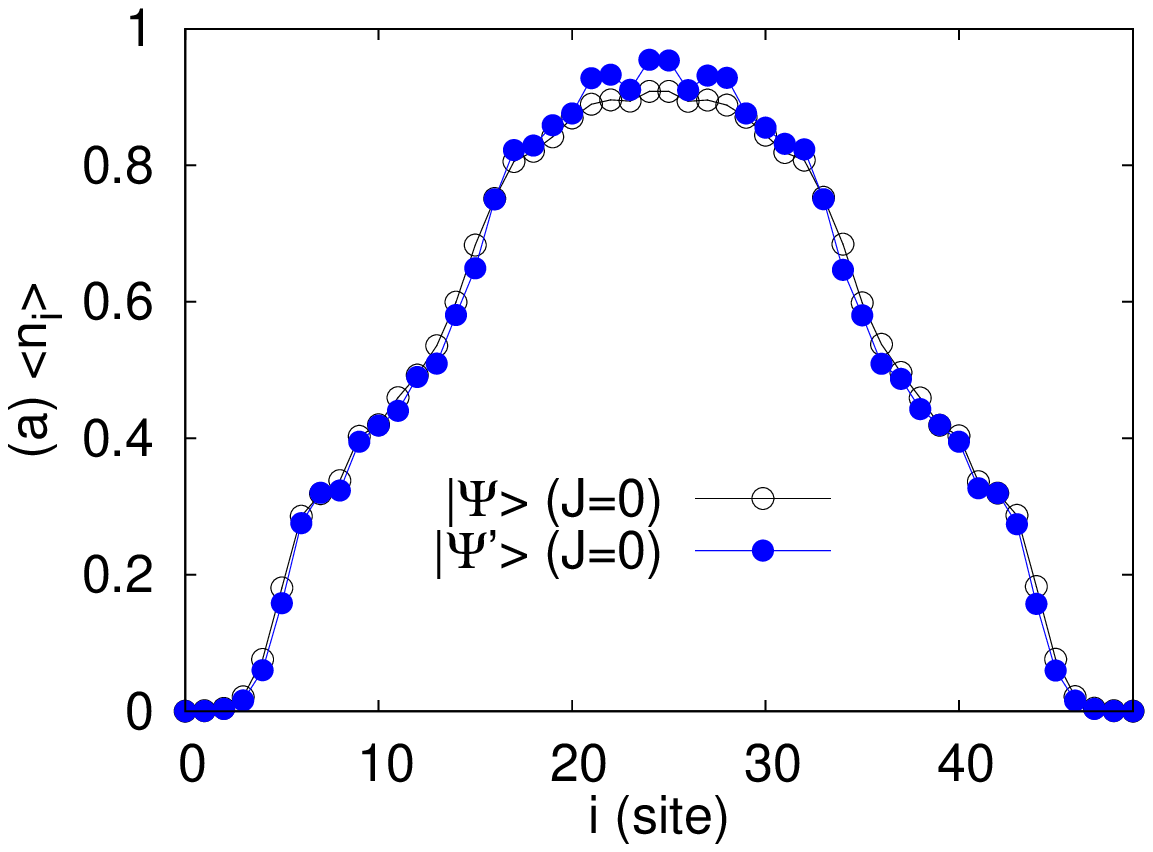}} &
\resizebox{42mm}{!}{\includegraphics{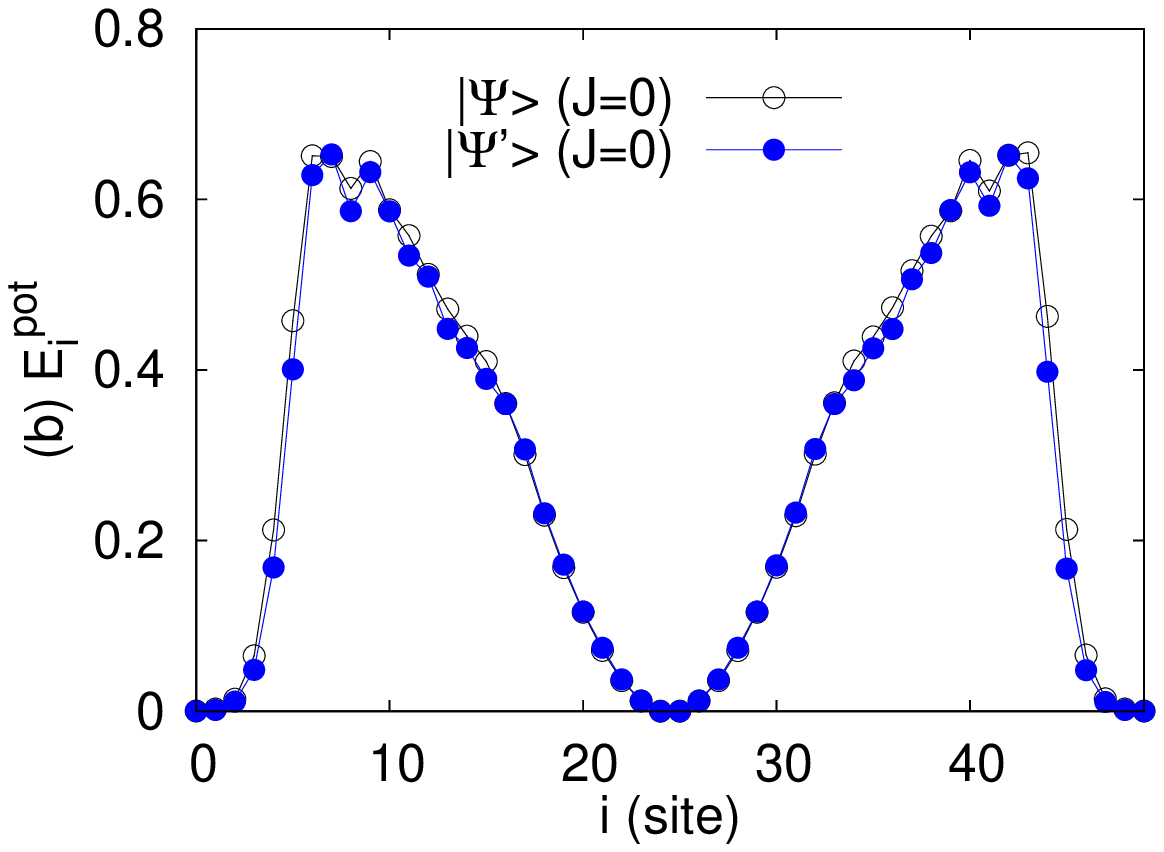}} \\
\hspace{-2mm}
\resizebox{42mm}{!}{\includegraphics{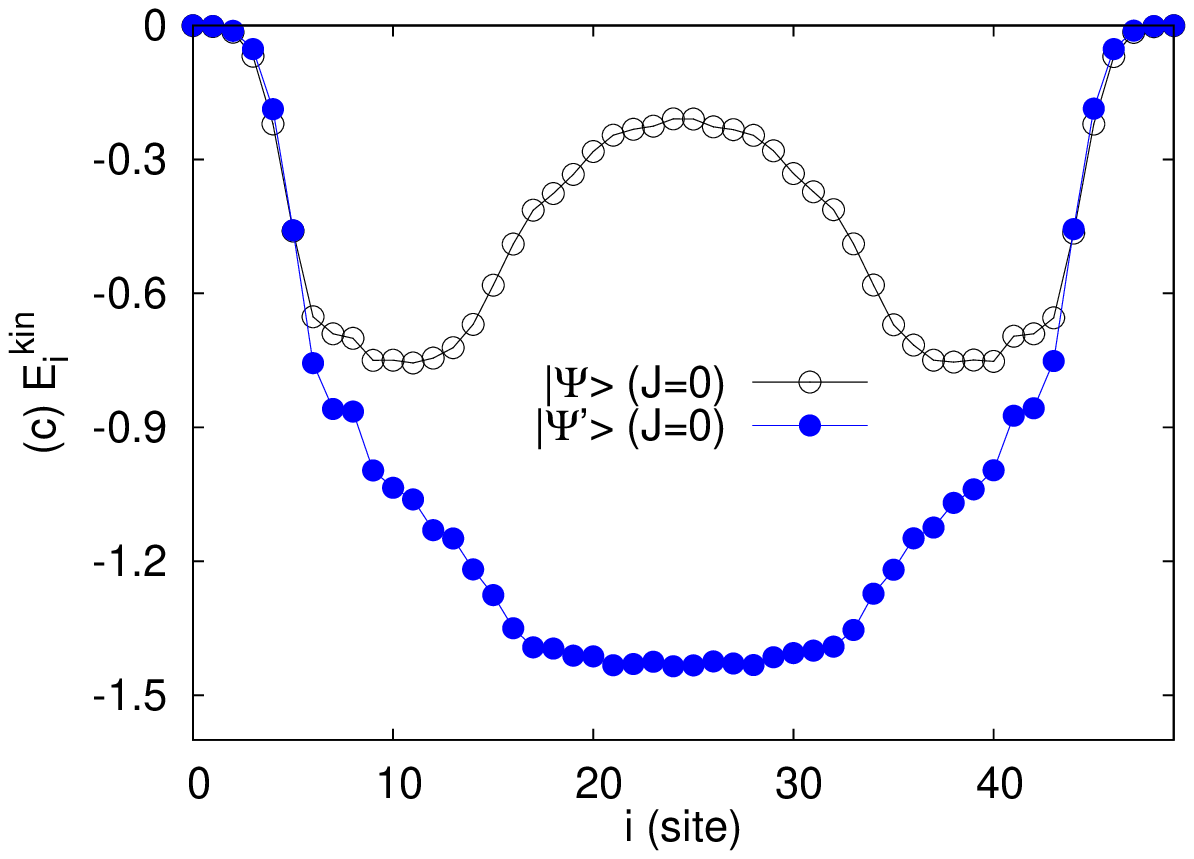}} &
\resizebox{42mm}{!}{\includegraphics{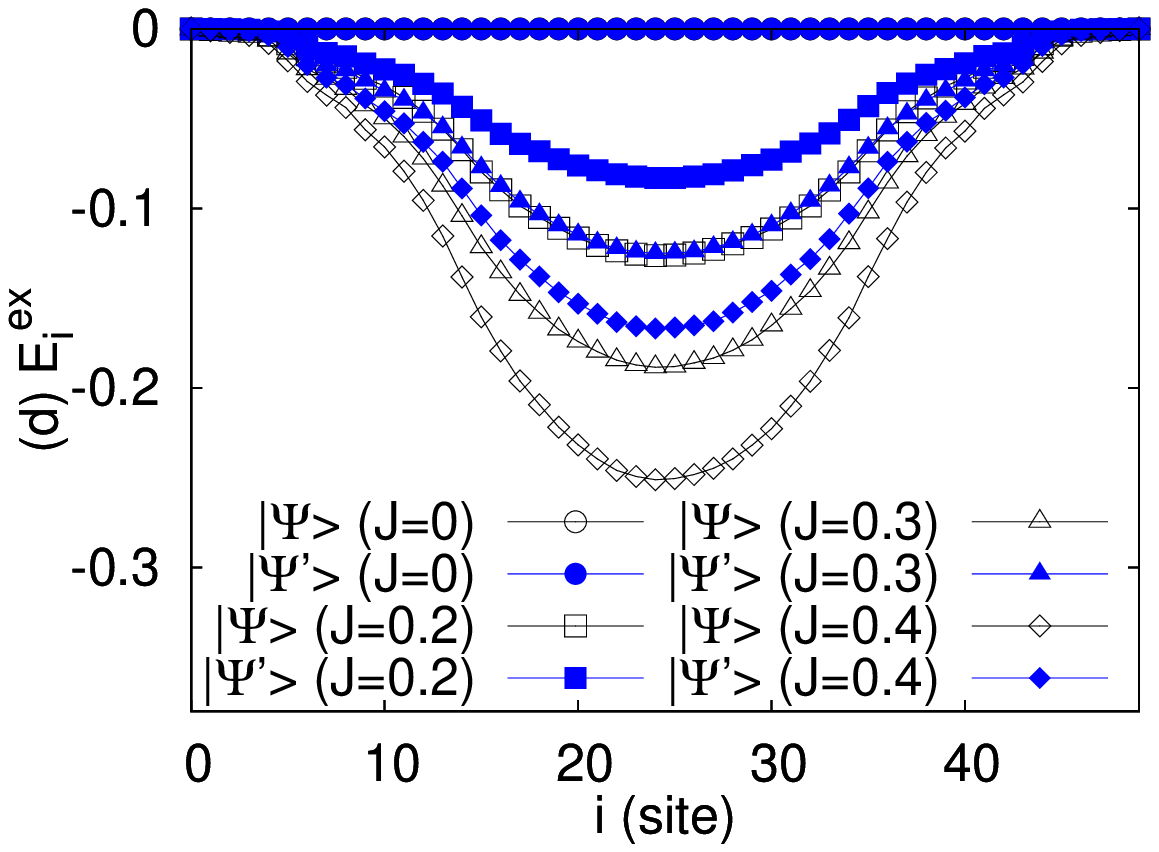}} \\
\end{tabular}
\caption{(Color online) (a) Particle density $\langle n_{i} \rangle$, and local average of energies: (b) potential energy $E_{i}^{pot}$, (c) kinetic energy $E_{i}^{kin}$ and (d) exchange energy $E_{i}^{ex}$ obtained with the Gutzwiller-projected state $| \Psi \rangle$ (open symbols) and the non-projected state $| \Psi' \rangle$ (filled symbols). We note that 
$\langle n_{i} \rangle$, $E_{i}^{pot}$ and $E_{i}^{kin}$ hardly depend on $J$.
}
\label{Fig_local}
\end{figure}

In order to elucidate the role of the onsite correlation effects from a microscopic point of view, we now compare the VMC results obtained with the Gutzwiller-projected state (\ref{TWF}) with those of $|\Psi' \rangle = |\Phi_{BCS} (\{ \Delta_{ij} \}, g) \rangle$. Since the Gutzwiller projection is neglected in the latter, strong onsite correlation effects are not incorporated in the results obtained with $|\Psi' \rangle$. Here we compute the local average of potential, kinetic and exchange energies, which are respectively defined by $E_{i}^{pot} \equiv V_{i} \langle n_{i} \rangle$,  $E_{i}^{kin} \equiv -t \sum_{j\sigma} \delta_{\langle ij \rangle} \langle c_{i\sigma}^{\dagger}c_{j\sigma} + h.c. \rangle$, $E_{i}^{ex} \equiv (J/2) \sum_{j} \delta_{\langle ij \rangle} \langle \mathbf{\hat{S}}_{i} \cdot \mathbf{\hat{S}}_{j}\rangle$. Shown in Fig.\ref{Fig_local} are the profiles of particle density and the local average of energies obtained by the VMC simulations. It is seen that the onsite correlation effects hardly change the density profile of particles (Fig.\ref{Fig_local} (a)). 
As a result, there is little difference in the potential energy $E_{i}^{pot}$ obtained with $|\Psi \rangle$ and $|\Psi' \rangle$, as seen in Fig.\ref{Fig_local} (b). On the other hand, the introduction of the onsite correlation effects significantly reduces $E_{i}^{kin}$ around the center of the system (Fig.\ref{Fig_local} (a)), because the strong repulsive interactions renormalize the motion of particles, forming heavy fermion states in the region with high particle density. In fact, the ratio of the two kinetic energies evaluated by Gutzwiller-projected and non-projected states can be regarded as the site-dependent wave-function renormalization factor $z_i$: $z_i \sim 1.0$ ($z_i \sim 0.2$) around the edge (center) of the system in Fig.\ref{Fig_local} (c). We note that these results hardly depend on $J$. On the other hand, it is seen in Fig.\ref{Fig_local} (d) that the antiferromagnetic spin correlations are developed among the heavy particles around the center of the system, where the energy gain due to $E_{i}^{ex}$ is enhanced in the presence of the strong correlation effects. Summarizing the above results, the onsite correlation effects enhance the energy gain by the exchange interactions, although they cause the loss of the kinetic energy. An important point is that in the high density region ($\langle n_{i} \rangle \sim 1.0$), the effective exchange coupling $J$ divided by the renormalized hopping $z_i t$ can be very large, which is efficient to stabilize the $d$-wave superfluid mediated by antiferromagnetic correlations. This naturally explains why the critical value $J\sim 0.2$ obtained for correlated systems can be smaller than $J \sim 0.5$ for non-interacting systems.

Note that the above conclusion holds true for the density regime with $n_{i} < 1.0$. For the system with  higher particle density, the spatial distribution of particles is expected to form a plateau around the center so as not to exceed $n_{i} = 1.0$, because the strong particle correlations have a tendency to form the Mott insulating state. Even in such cases, the present conclusion may be applied to the strongly correlated conducting region in the vicinity of  the Mott insulating state.

\section{Summary}

We have investigated the \textit{t-J} model to discuss the stability of the spatially-modulated superfluid state in a fermionic optical ladder system with harmonic confinement. By means of the BdG equations, we have clarified that the spatially-modulated superfluid state with $d_{x^{2}-y^{2}}$-like symmetry can be realized via antiferromagnetic exchange interactions. It has been found that the spatial distribution of the pair potential has a characteristic profile with cusps and plateau, which reflects the local density of states integrated around the Fermi level. Furthermore, the VMC calculation of the condensation energy as well as   the local average of potential, kinetic and exchange energies has shown that the local particle correlations enhance the stability of the superfluid state through energy gain due to singlet pairing in the region of high density filling ($\langle n_{i} \rangle \sim 1.0$).  In particular, we have demonstrated that the strong renormalization of kinetic energy around the high density region plays an important role in stabilizing the singlet-pairing superfluid state. We think that this kind of behavior may be characteristic of the systems with nonuniform distributions of the particle density.

In this paper, we have exploited the {\it t-J} model as an effective model in the strong coupling regime. In this model the exchange interaction $J$ is assumed to be free parameters, in contrast to the strong-coupling limit of the original Hubbard model. It thus remains an important problem to discuss the superfluidity by directly dealing with the Hubbard ladder system with a trapping potential, which is now in progress. 

When the density of particles is increased, other states such as the Mott insulating state and the antiferromagnetically ordered state may appear around the center of the system \cite{Fujihara, Rigol, Trebst, Andersen, Helmes}. Therefore, it is particularly worth studying how the $d$-wave like superfluid state coexists or competes with the magnetically ordered state in the optical lattice system, which is to be addressed in the future study.

\section*{Acknowledgments}
The numerical computations were carried out at the Supercomputer Center, the Institute for Solid State Physics, University of Tokyo.
This work is supported by Grant-in-Aids for Scientific Research [Grant nos. 20740194 (A.K.), and 19014013, 20029013 (N.K.)], and the Global COE Program "The Next Generation of Physics, Spun from Universality and Emergence" from the Ministry of Education, Culture, Sports, Science and Technology (MEXT) of Japan. 
Y. F is supported by JSPS Research Fellowships for Young Scientists.


\begin{thebibliography}{99}
\bibitem{Anderson}
M. H. Anderson, J. R. Ensher, M. R. Matthews, C. E. Wieman, and E. A. Cornell,
Science \textbf{269}, 198 (1995).

\bibitem{Bradley}
C. C. Bradley, C. A. Sackett, J. J. Tollett, and R. G. Hulet,
Rhys. Rev. Lett. \textbf{75}, 1687 (1995).

\bibitem{Davis}
K. B. Davis, M. -O. Mewes, M. R. Andrews, N. J. van Druten, D. S. Durfee, D. M. Kurn, and W. Ketterle,
Rhys. Rev. Lett. \textbf{75}, 3969 (1995).


\bibitem{Greiner}
M. Greiner, O. Mandel, T, Esslinger, T. W. H{\"a}nsch, and I. Bloch,
Nature \textbf{415}, 39 (2002).

\bibitem{Stoferle}
T. St{\"o}ferle, H. Moritz, C. Schori, M. K{\"o}hl, and T. Esslinger,
Phys. Rev. Lett. \textbf{92}, 130403 (2004).

\bibitem{Folling}
S. F{\"o}lling, A. Widera, T. M{\"u}ller, F. Gerbier, and I. Bloch,
Phys. Rev. Lett. \textbf{97}, 060403 (2006).

\bibitem{Roati}
G. Roati, E. de Mirandes, F. Ferlaino, H. Ott, G. Modugno and M. Inguscio,
Phys. Rev. Lett. \textbf{92} (2004) 230402.

\bibitem{Stoferle2}
T. St{\"o}ferle, H. Moritz, K. G{\"u}nter, M. K{\"o}hl and T. Esslinger,
Phys. Rev. Lett. \textbf{96} (2006) 030401.

\bibitem{Chin}
J. K. Chin, D. E. Miller, Y. Liu, C. Stan, W. Setiawan, C. Sanner, K. Xu, and W. Ketterle,
Nature \textbf{443}, 961 (2006).

\bibitem{Jordens}
R. J{\"o}rdens, N. Strohmaier, K. G{\"u}nter, H. Moritz, and T. Esslinger,
cond-mat/0804.4009.


\bibitem{Kohl}
M. K{\"o}hl, H. Moritz, T. St{\"o}ferle, K. G{\"u}nter, and T. Esslinger,
Phys. Rev. Lett. \textbf{94}, 080403 (2005).


\bibitem{Koetsier}
A. Koetsier, D. B. M. Dickerscheid, and H. T. C. Stoof,
Phys. Rev. A \textbf{74}, 033621 (2006).

\bibitem{Zhao}
E. Zhao, and A. Paramekanti,
Phys. Rev. Lett. \textbf{97}, 230404 (2006).

\bibitem{Tamaki}
H. Tamaki, Y. Ohashi, and K. Miyake,
Phys. Rev. A \textbf{77}, 063616 (2008).


\bibitem{Orso}
G. Orso, and G. V. Shlyapnikov,
Phys. Rev. Lett. \textbf{95}, 260402 (2005).

\bibitem{Zhai}
H. Zhai, and T. L. Ho,
Phys. Rev. Lett. \textbf{99}, 100402 (2007).

\bibitem{Moon}
E. G. Moon, P. Nikolic, and S. Sachdev,
Phys. Rev. Lett. \textbf{99}, 230403 (2007).

\bibitem{Iskin2}
M. Iskin, and C. A. R. S{\'a} de Melo,
Phys. Rev. A \textbf{78}, 013607 (2008).

\bibitem{Chien}
C. C. Chien, Y. He, Q. Chen, and K. Levin,
Phys. Rev. A \textbf{77}, 011601(R) (2008).


\bibitem{Pour}
F. K. Pour, M. Rigol, S. Wessel, and A. Muramatsu,
Rhys. Rev. B \text{75}, 161104 (2007).

\bibitem{Xianlong}
G. Xianlong, M. Rizzi, Marco Polini, R. Fazio, M. P. Tosi, V. L. Campo, Jr., and K. Capelle,
Rhys. Rev. Lett. \text{98}, 030404 (2007).

\bibitem{Dao}
T. -L. Dao, A. Georges, and M. Capone,
Rhys. Rev. B \text{76}, 104517 (2007).

\bibitem{Koga}
A. Koga, T. Higashiyama, K. Inaba, S. Suga, and N. Kawakami,
J. Phys. Soc. Jpn. \textbf{77}, 073602 (2008).

\bibitem{Burkov}
A. A. Burkov, and A. Paramekanti,
cond-mat/0802.2101.


\bibitem{Moreo}
A. Moreo, and D. J. Scalapino,
Phys. Rev. Lett. \textbf{98}, 216402 (2007).

\bibitem{Parish}
M. M. Parish, S. K. Baur, E. J. Mueller, and D. A. Huse,
Phys. Rev. Lett. \textbf{99}, 250403 (2007).

\bibitem{Chen}
Y. Chen, Z. D. Wang, F. C. Zhang, and C. S. Ting,
cond-mat/0710.5484.

\bibitem{Iskin}
M. Iskin, and C. J. Williams,
Phys. Rev. A \textbf{78}, 011603(R) (2008).

\bibitem{Rizzi}
M. Rizzi, M. Polini, M. A. Cazalilla, M. R. Bakhtiari, M. P. Tosi, and R. Fazio,Phys. Rev. B \textbf{77}, 245105 (2008).

\bibitem{Koponen}
T. K. Koponen, T. Paananen, J. -P. Martikainen, M. R. Bakhtiari, and P. T{\"o}rm{\"a},
New. J. Phys. \textbf{10}, 045014 (2008).


\bibitem{Bloch2}
I. Bloch,
Nature Physics. \textbf{1}, 23 (2005).

\bibitem{Jaksch2}
D. Jaksch, and P. Zoller,
Ann. Phys. \textbf{315}, 52 (2005).

\bibitem{Morsch}
O. Morsch, and M. Oberthaler,
Rev. Mod. Phys. \textbf{78}, 179 (2006).

\bibitem{Bloch}
I. Bloch, J. Dalibard, and W. Zwerger,
Rev. Mod. Phys. \textbf{80}, 885 (2008).


\bibitem{Bednorz}
J. G. Bednorz, and K. A. M{\"u}ller,
Z. Phys. B \textbf{64}, 189 (1986).

\bibitem{Yanase}
Y. Yanase, T. Jujo, T. Nomura, H. Ikeda, T. Hotta, and K. Yamada,
Phys. Rep. \textbf{387}, 1 (2003), and reference therein.

\bibitem{Levin}
Q. J. Chen, C. -C. Chien, Y. He, and K. Levin,
J. Supercond. Nov. Magn. \textbf{20}, 515 (2007).

\bibitem{Akimitsu}
M. Uehara, T. Nagata, J. Akimitsu, H. Takahashi, N. M{\^o}ri, and K. Konshita,
J. Phys. Soc. Jpn. \textbf{65}, 2764 (1996).


\bibitem{Yokoyama}
H. Yokoyama, and H. Shiba,
J. Phys. Soc. Jpn. \textbf{56}, 1490 (1987).

\bibitem{Fujihara}
Y. Fujihara, A. Koga, and N. Kawakami,
J. Phys. Soc. Jpn. \textbf{76}, 034716 (2007):
J. Magn. Magn. Mater. \textbf{310}, 882 (2007):
J. Phys. Chem. Solids. \textit{in press.}


\bibitem{Fukuhara}
T. Fukuhara, Y. Takasu, M. Kumakura, and Y. Takahashi,
Phys. Rev. Lett. \textbf{98}, 030401 (2007).

\bibitem{Jaksch}
D. Jaksch, C. Bruder, J. I. Cirac, C. W. Gardiner, and P. Zoller,
Phys. Rev. Lett. \textbf{81}, 3108 (1998).

\bibitem{Hofstetter}
W. Hofstetter, J. I. Cirac, P. Zoller, E. Demler, and M. D. Lukin,
Phys. Rev. Lett. \textbf{89}, 220407 (2002).

\bibitem{Strabley}
J. Sebby-Strabley, M. Anderlini, P. S. Jessen, and J. V. Porto
Phys. Rev. A \textbf{73}, 033605 (2006),

\bibitem{Danshita}
I. Danshita, J. E. Williams, C. A. R. S{\'a} de Melo, and C. W. Clark,
Phys. Rev. A \textbf{76}, 043606 (2007)


\bibitem{Duan}
L. -M. Duan, E. Demler, and M. D. Lukin,
Phys. Rev. Lett. \textbf{91}, 090402 (2003).

\bibitem{Kuklov}
A. B. Kuklov, and B. V. Svistunov,
Phys. Rev. Lett. \textbf{90}, 100401 (2003).

\bibitem{Ripoll}
J. J. Garc{\'i}a-Ripoll, and J. I. Cirac,
New. J. Phys. \textbf{5}, 76 (2003).

\bibitem{Trotzky}
S. Trotzky, P. Cheinet, S. F{\"o}lling, M. Feld, U. Schnorrberger, A. M. Rey, A. Polkovnikov, E. A. Demler, M. D. Lukin, and I. Bloch,
Science \textbf{20}, 1150841 (2007).


\bibitem{Ogata}
M. Ogata,
J. Phys. Soc. Jpn. \textbf{66}, 3375 (1997).


\bibitem{Sigrist}
M. Sigrist, T. M. Rice, and F. C. Zhang,
Phys. Rev. B \textbf{49}, 12058 (1994).

\bibitem{Troyer}
M. Troyer, H. Tsunetsugu, and T. M. Rice,
Phys. Rev. B \textbf{53}, 251 (1996).

\bibitem{Orignac}
E. Orignac, and T. Giamarchi,
Phys. Rev. B, \textbf{56}, 7167 (1997).

\bibitem{Sorella}
S. Sorella, G. B. Martins, F. Becca, C. Gazza, L. Capriotti, A. Parola, and E. Dagotto,
Phys. Rev. Lett. \textbf{88}, 117002 (2002).


\bibitem{Zhang}
F. C. Zhang, C. Gros, T. M. Rice, and H. Shiba,
Supercond. Sci. Technol. \textbf{1}, 36 (1988).

\bibitem{Ogata2}
H. Yokoyama, and M. Ogata,
J. Phys. Soc. Jpn. \textbf{65}, 3615 (1996).

\bibitem{Ogata3}
A. Himeda, M. Ogata, Y. Tanaka, and S. Kashiwaya,
J. Phys. Soc. Jpn. \textbf{66}, 3367 (1997).




\bibitem{Gutzwiller}
M. C. Gutzwiller,
Phys. Rev. Lett. \textbf{10}, 159 (1963).


\bibitem{Rigol}
M. Rigol, A. Muramatsu, G. G. Batrouni, R. T. Scalettar,
Phys. Rev. Lett. \textbf{91}, 130403 (2003).

\bibitem{Trebst}
S. Trebst, U. Schollw{\"o}ck, M. Troyer, and P. Zoller,
Phys. Rev. Lett. \textbf{96}, 250402 (2006).

\bibitem{Andersen}
B. M. Andersen, and G. M. Bruun,
Phys. Rev. A \textbf{76}, 041602(R) (2007).

\bibitem{Helmes}
R. W. Helmes, T. A. Costi, and A. Rosch,
Phys. Rev. Lett. \textbf{100}, 056403 (2008).


\end{thebibliography}
\end{document}